\documentclass[aps,prx,twocolumn,superscriptaddress,showpacs]{revtex4-1}
\usepackage{hyperref}
\usepackage{times,xspace}
\usepackage{graphicx,graphics,color,epsfig}
\usepackage{amsmath}
\usepackage{amssymb}
\usepackage{amsfonts}
\usepackage{amsfonts}
\usepackage{epstopdf}
\usepackage{bm}
\usepackage{longtable}     
\usepackage{url}           
\usepackage{bm}            
\usepackage{color}
\usepackage{float}
\usepackage[normalem]{ulem}

\newcommand{\ourtitle}{Superconductivity in Infinite-layer Nickelates : Role of Non-zero $f$-ness}

\begin{document}

\title{\ourtitle}
\author{Subhadeep Bandyopadhyay$^{\ast}$}
\affiliation{School of Physical Sciences, Indian Association for the Cultivation of Science, Kolkata 700 032, India.}
\author{Priyo Adhikary$^{\ast}$}
\affiliation{Department of Physics, Indian Institute of Science, Bangalore 560012, India.}
\author{Tanmoy Das}
\email{tnmydas@gmail.com}
\affiliation{Department of Physics, Indian Institute of Science, Bangalore 560012, India.}
\author{Indra Dasgupta}
\email{sspid@iacs.res.in}
\affiliation{School of Physical Sciences, Indian Association for the Cultivation of Science, Kolkata 700 032, India.}
\author{Tanusri Saha-Dasgupta}
\email{t.sahadasgupta@gmail.com}
\affiliation{S. N. Bose National Centre for Basic Sciences, JD Block, Sector III, Salt Lake, Kolkata, West Bengal 700106, India.}

\begin{abstract}
Employing first-principles density functional theory calculations and Wannierization of
the low energy band structure, we analyze the electronic structure of undoped, infinite-layer nickelate compounds, NdNiO$_2$, PrNiO$_2$ and LaNiO$_2$. Our study reveals important role of non-zero $f$-ness of Nd and Pr atoms, as opposed to $f^{0}$ occupancy of La. The non-zero $f$-ness becomes effective in lowering the energy of the rare-earth 5$d$ hybridized {\it axial orbital} thereby enhancing the electron pockets and influencing the Fermi surface topology. The Fermi surface topology of NdNiO$_2$ and PrNiO$_2$ is strikingly similar, while differences are observed for LaNiO$_2$. This difference shows up in computed doping dependent superconducting properties of the three compounds within a weak coupling theory, which finds two gap superconductivity for NdNiO$_2$ and PrNiO$_2$, and  possibility of a single gap superconductivity for LaNiO$_2$ with the strength of superconductivity suppressed by almost a factor of two, compared to Nd or Pr compound.
 
\end{abstract}  
\maketitle

{\it Introduction.--} Discovery of high temperature superconductivity in cuprates\cite{cuprates} consisting of two-dimensional (2D) CuO$_2$ planes, has prompted search for other transition metal oxide compounds, having similar structural motif as in high T$_c$ cuprates. This raised the interest in the existence of nickelate compounds with 2D NiO$_2$ planes as the common structural element, having Ni$^{1+}$ ions of electronic
configuration $d^9$, isoelectronic to Cu$^{2+}$. 

The first nickelate structure containing 2D NiO$_2$ planes that was synthesized is the infinite layer LaNiO$_2$.\cite{lno} Subsequently, another member of the infinite layer nickelate series, namely, NdNiO$_2$ has been synthesized.\cite{nno} Very recently, NdNiO$_2$ has been shown\cite{nno-sc} to be superconducting upon hole doping with T$_c$ $\approx$ 9-15 K with a number of studies devoted on this topic.\cite{aoki,rony,millis-nno,karsten,JiangDFT,GuDFT,AritaDFT,pickett} Superconductivity in LaNiO$_2$ is not yet reported, though NdNiO$_2$ and LaNiO$_2$ are isostructural. This makes the origin of possible differential behavior of LaNiO$_2$ and NdNiO$_2$ intriguing. The ionic radius\cite{ionic} of La$^{3+}$ is 104 pm $\approx$ 4\% larger than the ionic radius of Nd$^{3+}$ (100 pm), resulting in an expanded lattice of volume $\approx$ 5\% in LaNiO$_2$ compared to NdNiO$_2$.\cite{nno,lno-struc} Such structural differences may influence the electronic behavior, though this has not been investigated. On the other hand, it has been suggested\cite{pickett19} that direct hybridization with the Nd 4$f$ states may become important for the description of electronic structure of NdNiO$_2$ near the chemical potential. It is to be noted that the spin disorder broadening induced by such direct hybridization is expected to play a detrimental role in superconductivity rather than helping it. The situation became further curious by the report of PrNiO$_2$ \cite{pno} exhibiting superconductivity with T$_c$ $\approx$ 7-12 K. Very similar values of T$_c$ for NdNiO$_2$ and PrNiO$_2$, despite Nd$^{3+}$ having a $f$-electron count of 3 and Pr$^{3+}$ having a $f$-electron count of 2, raises questions on the active role of $f$-electrons on the electronic behaviour and consequent superconducting properties of NdNiO$_2$ and PrNiO$_2$. This leaves the role of 4$f$ electrons an open issue. On the other hand, it is interesting to note that while the ground state for the quasi-2D reduced trilayer nickelate La$_4$Ni$_3$O$_8$ is a charge density-wave insulator,\cite{lanio} that of Pr$_4$Ni$_3$O$_8$\cite{prnio} is metallic, implying the difference between the La and Pr systems manifests not only in case of infinite layer 2D compounds, but also for quasi-2D compounds.

In this communication, we probe this issue by computing the electronic structure of NdNiO$_2$, PrNiO$_2$ and LaNiO$_2$ within the framework of density functional theory, probing the impact of structural and chemical changes on the electronic structure, on the active orbitals participating on the formation of Fermi surface and on superconductivity. The comparison and contrast between the three compounds, two of them with reported superconducting properties,\cite{nno-sc,pno}
and one not-yet-reported to be superconducting, unravels the possible role of non-zero $f$-ness in this curious trend.

\begin{figure}
\begin{center}
\rotatebox{0}{\includegraphics[width=0.5\textwidth]{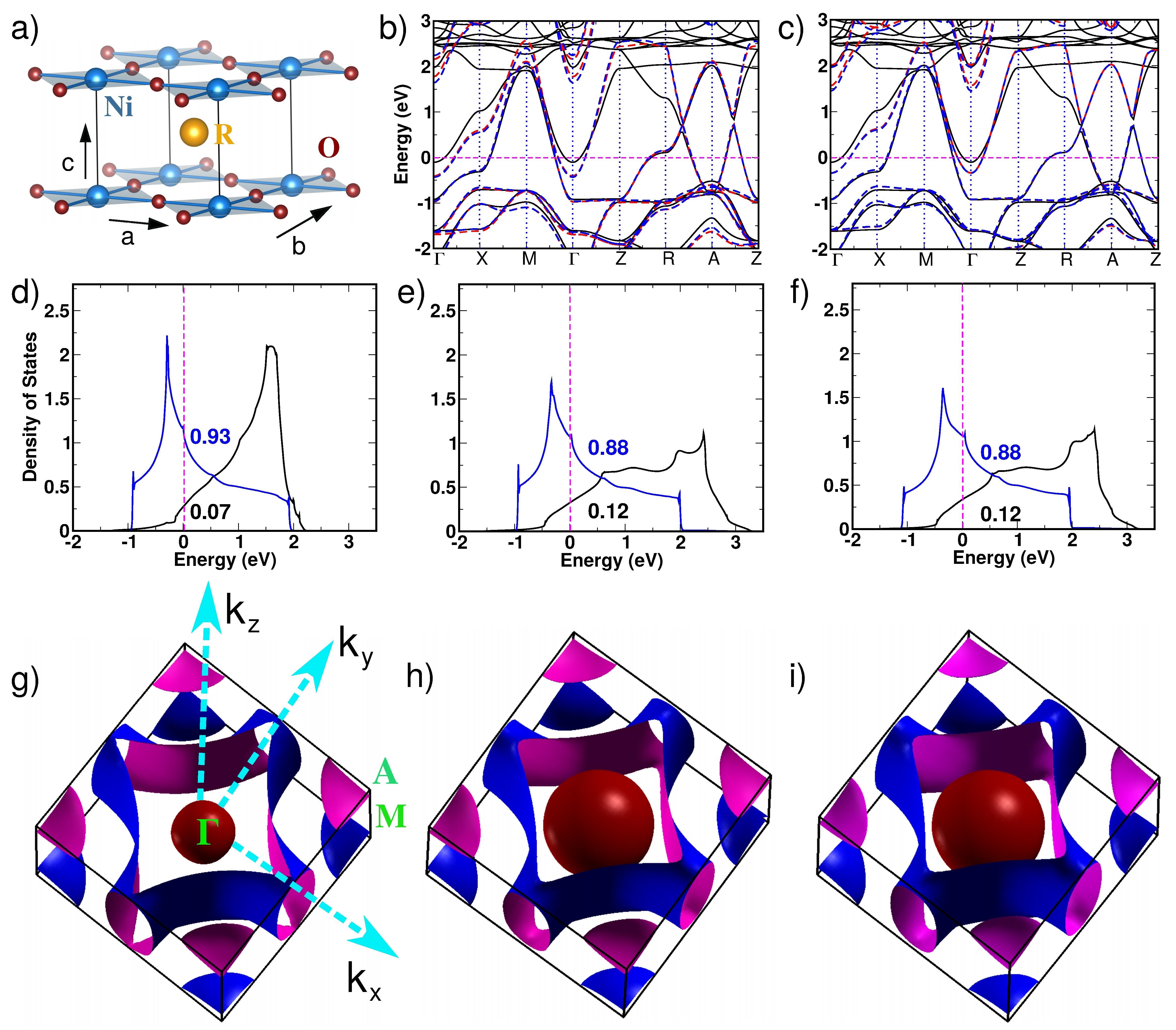}}
\end{center}
\caption{(Color online) a) Crystal structure of infinite-layer RNiO$_2$ (R = Pr/Nd/La). b) DFT band structure of LaNiO$_2$ (black, solid lines) in comparison to NdNiO$_2$ (red, dashed lines) and PrNiO$_2$ (blue, dashed lines) plotted along the high symmetry lines ($\Gamma$(0,0,0)-X($\pi$/a,0,0)-M($\pi$/a,$\pi$/a,0)-$\Gamma$-Z(0,0,$\pi$/c)-R($\pi$/a,0,$\pi$/c)-A($\pi$/a,$\pi$/a,$\pi$/c) of the tetragonal Brillouin zone, computed for the actual crystals of LaNiO$_2$, NdNiO$_2$ and PrNiO$_2$. c) Same as b) but band structures computed using the same crystal
  structure (crystal structure of LaNiO$_2$) for all three compounds.
 The zero of the energy is set Fermi level. d)-f) The density of states corresponding
  to two-bands crossing Fermi level for LaNiO$_2$ (d), PrNiO$_2$ (e) and NdNiO$_2$ (f). Labeled are the occupancies of the two bands.
  g)-i) The Fermi surfaces for LaNiO$_2$ (g), PrNiO$_2$ (h) and NdNiO$_2$ (i). 
}
\label{fig1}
\end{figure}

{\it Electronic Structure.--} Fig. 1a) shows the infinite-layer structure of nickelate compounds with general formula RNiO$_2$ (R = La, Nd, Pr). The La/Pr/Nd atoms occupy the position (0.5,0.5,0.5), with Ni at (0,0,0) and O at (0.5,0.0,0.0) of the P4/mmm space group. The lattice constants of the tetragonal unit cell,\cite{nno,lno-struc,pno} are given by, $a$ = $b$ = 3.92/3.91/3.96 \AA, and $c$ =  3.28/3.31/3.36 for Nd/Pr/La compounds.  The comparison of the electronic structure of the three compounds, in their respective crystal structures and that in the same crystal structure as that of LaNiO$_2$, as computed within density functional theory (DFT), are shown in Fig.s 1b) and 1c), respectively. DFT calculations have been carried out in plane wave basis with projected augmented wave (PAW) potential,\cite{PAW} as implemented in Vienna Ab-initio Simulation Package (VASP)\cite{vasp} with choice of generalized gradient approximation (GGA)\cite{PBE} for the exchange-correlation functional.\cite{suppl} Although the band structure of the three compounds look similar at a first glance, with two low-energy bands crossing Fermi level giving rise to a hole pocket centered
around M(A) point, and electron pockets at $\Gamma$ and A points, there are interesting and important differences.
First of all, we find while the band structures of Nd and Pr compounds fall almost on top of each other, there exists significant differences with the band structure of LaNiO$_2$ (cf Fig. 1b). In particular, there is a marked difference observed in the electron pocket around $\Gamma$-point which arise due to hybridization between R-5$d$ ($3z^{2}$-$r^{2}$) states and Ni-3$d$ states.  For the band structures obtained with fixed crystal structure (panel c)) the electron pocket around $\Gamma$-point in Nd/PrNiO$_2$ lies about 0.2 eV below that in LaNiO$_2$. The structural differences between Nd/PrNiO$_2$ and LaNiO$_2$, pushes the electron pocket around $\Gamma$-point further down in Nd/PrNiO$_2$, making it lie about 0.4 eV below that in LaNiO$_2$ (panel b)). This difference in electronic structure between Nd/PrNiO$_2$ and LaNiO$_2$ can be appreciated further in terms of the density of states corresponding to two low-energy bands crossing the Fermi level. We find a marked difference in the occupancies of the two bands between Nd/PrNiO$_2$ and LaNiO$_2$, being 0.88 e$^{-}$ and 0.12 e$^{-}$ for Nd/PrNiO$_2$ in contrast to 0.93 e$^{-}$ and 0.07 e$^{-}$ for LaNiO$_2$. This brings in significant difference in fermiology between Nd/PrNiO$_2$ and LaNiO$_2$ with $\Gamma$-centered spherical hole pocket in Nd/PrNiO$_2$ being about twice the size of that in LaNiO$_2$.

\begin{figure}
\begin{center}
\rotatebox{0}{\includegraphics[width=0.5\textwidth]{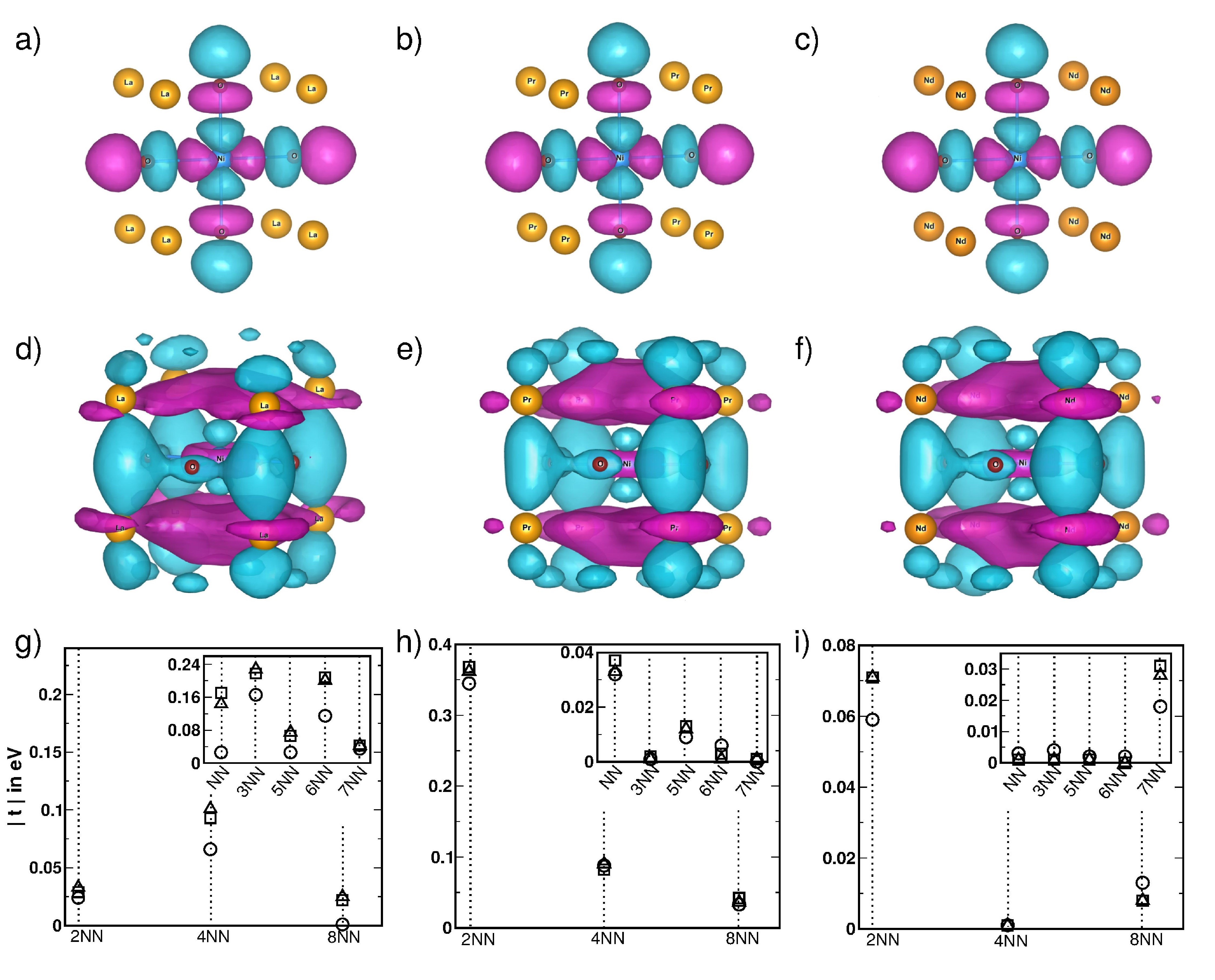}}
\end{center}
\caption{(Color online) a)-c) The effective Ni $x^{2}$-$y^2$ Wannier function for LaNiO$_2$ (a), PrNiO$_2$ (b) and NdNiO$_2$ (c). The oppositely signed lobes are colored differently. d)-f) The effective axial Wannier function for LaNiO$_2$ (c), PrNiO$_2$ (d) and NdNiO$_2$ (e). g)-i) The in-plane and out-of-plane (insets) intra- and inter-orbital
  hopping interactions plotted as function of Ni-Ni distances: between axial-axial (g), $x^{2}$-$y^2$-$x^{2}$-$y^2$ (h) and $x^{2}$-$y^2$-axial (i). The distances corresponding to various near-neighbors (NNs) are marked. The interactions for
  LaNiO$_2$, PrNiO$_2$ and NdNiO$_2$ are shown as circle, triangle and squares.}
\label{fig2}
\end{figure}

As shown previously,\cite{our} two bands crossing the Fermi-level can be spanned by two effective orbitals, Ni-$x^2$-$y^{2}$ orbital and an {\it axial} orbital formed by linear combination of R-$3z^2$-$r^{2}$, R-$xy$, Ni-$3z^{2}$-$r^2$ and
Ni-$s$, constructed using the formulation of WANNIER90.\cite{WAN}
The axial orbital formed is not centered on a single atom, rather its density encompasses both Ni and R atoms, with significant weight occupying the interstitial.\cite{AritaDFT} The effective $x^2$-$y^2$ and axial orbitals for the three compounds, NdNiO$_2$ (NNO), PrNiO$_2$ (PNO) and LaNiO$_2$ (LNO) are shown in Fig. 2a)-f). The effective $x^2$-$y^2$ orbital which forms an antibonding combination of Ni $x^2$-$y^2$ and O p$_x$/p$_y$ is identical in three compounds. On the other hand, while the features of axial orbital is same between NdNiO$_2$ and PrNiO$_2$, it is different in LaNiO$_2$. As found previously,\cite{our} it is thus the axial orbital that encodes the materials dependence. 
This difference in orbital characters spanning the low energy bands gets reflected in the intra-orbital and inter-orbital hopping interactions, shown in panels g)-i) of Fig. 2. In general, we find that hopping interactions in PrNiO$_2$ are similar to that of NdNiO$_2$, while that of LaNiO$_2$ is different. Most importantly, the inter-orbital hopping, connecting $x^2$-$y^2$ and the axial orbital is markedly different between Nd/PrNiO$_2$ and LaNiO$_2$, the hopping along [100] (2nd near-neighbor, 2NN) and [102] (7NN) in Nd/PrNiO2 being 17\% and  50\% larger than that of LaNiO$_2$. The remarkable similarity in the structure of axial orbital between NdNiO$_2$ and PrNiO$_2$ compounds and dissimilarity with LaNiO$_2$ highlights the fact that  NdNiO$_2$ and PrNiO$_2$ belongs to same class, while LaNiO$_2$ is different.  What causes this difference? The electronic configuration of La$^{3+}$ is [Xe] while that of Nd$^{3+}$ and Pr$^{3+}$ are  [Xe]4$f^3$ and [Xe]4$f^2$, respectively. This makes 4$f$ bands of La$^{3+}$ lie empty around 2.5 eV above the Fermi level, while three and two electrons of 4$f$ of Nd3+ and Pr3+ contribute in the core within the nonmagnetic scheme of calculation. The resultant effect of empty 4$f$ and 5$d$ interaction of La in case of LaNiO$_2$, vis-a-vis 4$f$ core for Nd/Pr puts the active 5$d$ ($3z^2$-$r^{2}$) state of Nd/PrNiO$_2$ in a favorable position to hybridize more effectively with the Ni states resulting in the band center of the hybrid axial orbital lying 0.5 eV lower in energy in comparison to LaNiO$_2$ and having bandwidth 1 eV larger in  Nd/PrNiO$_2$ compared to LaNiO$_2$. We note this effect arises irrespective of the disordered magnetism of 4$f$,\cite{pickett19} the role of which till date is not known.  This emphasizes the importance of the non-zero $f$-ness in Nd/Pr in tailoring the position of the axial orbital and consequent impact on the superconductivity which we discuss next. 

{\it Superconductivity:-} The natural question is how does this materials specific electronic structure influence the superconducting (SC) state in this family. We extend our previous studies on spin-fluctuation mediated SC pairing symmetry and pairing strength to all three materials and as a function of doping.\cite{our} Earlier, we have shown\cite{our} that the inter-orbital interaction between the Ni-$x^2$-$y^2$ and axial orbital plays the key role to stabilize SC ground state in these materials. The importance of inter-orbital interaction is supported by the spectroscopic data\cite{Hepting} as well as found in a recent DFT+DMFT study\cite{DMFT} on Nd$_{1-x}$Sr$_x$NiO$_2$. 


We consider a two-band Hubbard model with interaction Hamiltonian given by,
\begin{equation}
H_{\rm int} = \sum_{i,\sigma}U_i n_{i\sigma}n_{i\bar{\sigma}} + V\sum_{i\ne j,\sigma\sigma'} n_{i\sigma}n_{j\sigma'}, 
\end{equation}
where $n_{i\sigma}$ is the number density of the $i^{\rm th}$ (=1,2) orbital, and spin $\sigma/\bar{\sigma}$=$\uparrow/\downarrow$. $U_i$ are the intra-orbital interactions, and $V$ is the inter-orbital interaction. 

For the above interaction, an effective SC potential $\Gamma$ can be obtained by summing over the bubble and ladder diagrams\cite{SCrepulsive,SCcuprates,SCpnictides,SCHF} as
\begin{eqnarray}
H_{\rm int} &\approx& \frac{1}{\Omega_{\rm BZ}^2}\sum_{\alpha\beta\gamma\delta}\sum_{{\bf k,k'},\sigma\sigma'} \Gamma_{\alpha\beta}^{\gamma\delta}({\bf q})\nonumber\\
&&\quad\times c_{\alpha \sigma}^{\dagger}({\bf k})c_{\beta\sigma'}^{\dagger}(-{\bf  k})c_{\gamma\sigma'}({\bf -k'})c_{\delta\sigma}({\bf k'}).
\label{Hintpair}
\end{eqnarray}
$c^{\dag}_{\alpha,\sigma}({\bf k})$ is the creation operator for the $\alpha$-orbital with spin $\sigma=\uparrow/\downarrow$ at the wave vector ${\bf k}$. $\Gamma$ is the effective pairing tensor which can be decomposed into the singlet and triplet states. For NNO and also in PNO, the axial orbital assisted inter-orbital interaction $V$ promotes an orbital-selective superconductivity, while the orbital selectivity is nearly lost in LNO due to the lack of sufficient contribution of axial orbital to the Fermi surface (FS).\cite{our} The orbital dependent pairing fields in the spin-singlet channels ($\sigma'= -\sigma$) can be defined from the above equation as
\begin{eqnarray}
\Delta_{\alpha\beta}({\bf k})&=& -\frac{1}{\Omega_{\rm BZ}}\sum_{\gamma\delta\sigma}\sum_{{\bf k'}}
\Gamma_{\alpha\beta}^{\gamma\delta}({\bf k,k'})\left\langle c_{\gamma\sigma}({\bf -k'})c_{\delta\bar{\sigma}}({\bf k'})\right\rangle,\nonumber\\
&=& -\frac{1}{\Omega_{\rm BZ}}\sum_{\gamma\delta\sigma}\sum_{\bf k'} \Gamma_{\alpha\beta}^{\gamma\delta}({\bf k,k'})\Delta_{\gamma\delta}({\bf k}) \nonumber\\
&&\times\sum_{\nu}\frac{\phi^{\nu}_{\gamma}({\bf -k'})\phi^{\nu}_{\delta} ({\bf k'})}{2\xi_{\nu}({\bf k'})}{\rm tanh}\left(\frac{\xi_{\nu}({\bf k'})}{2k_BT}\right).
\end{eqnarray}
The expectation value on the right hand side is taken over the typical BCS ground state, which yields a self-consistent BCS gap equation in the second line. Here $\nu$ is the band index, and $\xi_{\nu}({\bf k})$ and $\phi^{\nu}_{\alpha}({\bf k})$ are the eigenvalue and eigenvector of the two-band Wannier Hamiltonian, directly obtained from the DFT code. At $T\rightarrow 0$ limit, and when the momenta are restricted to the FS, the ${\rm tanh}$-function leads to a $\delta$ function. Therefore, we introduce the SC coupling constant $\lambda\rightarrow \frac{1}{2\xi_{\nu}({\bf k})}{\rm tanh}\left(\frac{\xi_{\nu}({\bf k})}{2k_BT}\right)_{{\bf k\in {\bf k}_F}}$. Finally, we define the band dependent SC gap $\Delta_{\nu}$ and pairing potential $\Gamma_{\nu\nu'}$ as $\Delta_{\nu}({\bf k})=\sum_{\alpha\beta}\Delta_{\alpha\beta}({\bf k})\phi^{\nu*}_{\alpha}({\bf k})\phi^{\nu*}_{\beta}(-{\bf k})$, and  $\Gamma'_{\nu\nu'}({\bf k,k'})=\sum_{\alpha\beta\gamma\delta} \Gamma_{\alpha\beta}^{\gamma\delta}({\bf q})\phi^{\nu\dagger}_{\alpha}({\bf k})\phi^{\nu\dagger}_{\beta}(-{\bf k})\phi^{\nu'}_{\gamma}({\bf -k'})\phi^{\nu'}_{\delta} ({\bf k'})$.\cite{note} This yields the self-consistent SC gap equation, given by 
\begin{eqnarray}
\Delta_{\nu}({\bf k})= -\lambda\frac{1}{\Omega_{\rm BZ}}\sum_{\nu',{\bf q}}\Gamma_{\nu\nu'}({\bf k,q})\Delta_{\nu'}({\bf k+q}).
\label{SC2}
\end{eqnarray}
This is an eigenvalue equation of the pairing potential $\Gamma_{\nu\nu'}({\bf q}={\bf k}-{\bf k^{\prime}})$ with eigenvalue $\lambda$ and eigenfunction $\Delta_{\nu}({\bf k})$. The ${\bf k}$-dependence of $\Delta_{\nu}({\bf k})$ dictates the pairing symmetry for a given eigenvalue. While there are many solutions (as many as the ${\bf k}$-grid), we consider only the highest eigenvalue since this pairing symmetry can be shown to have the lowest Free energy value in the SC state.\cite{SCrepulsive} Further details on our computational method can be obtained in Ref.~\cite{our,SRay}

\begin{figure}
\begin{center}
\rotatebox{0}{\includegraphics[width=0.45\textwidth]{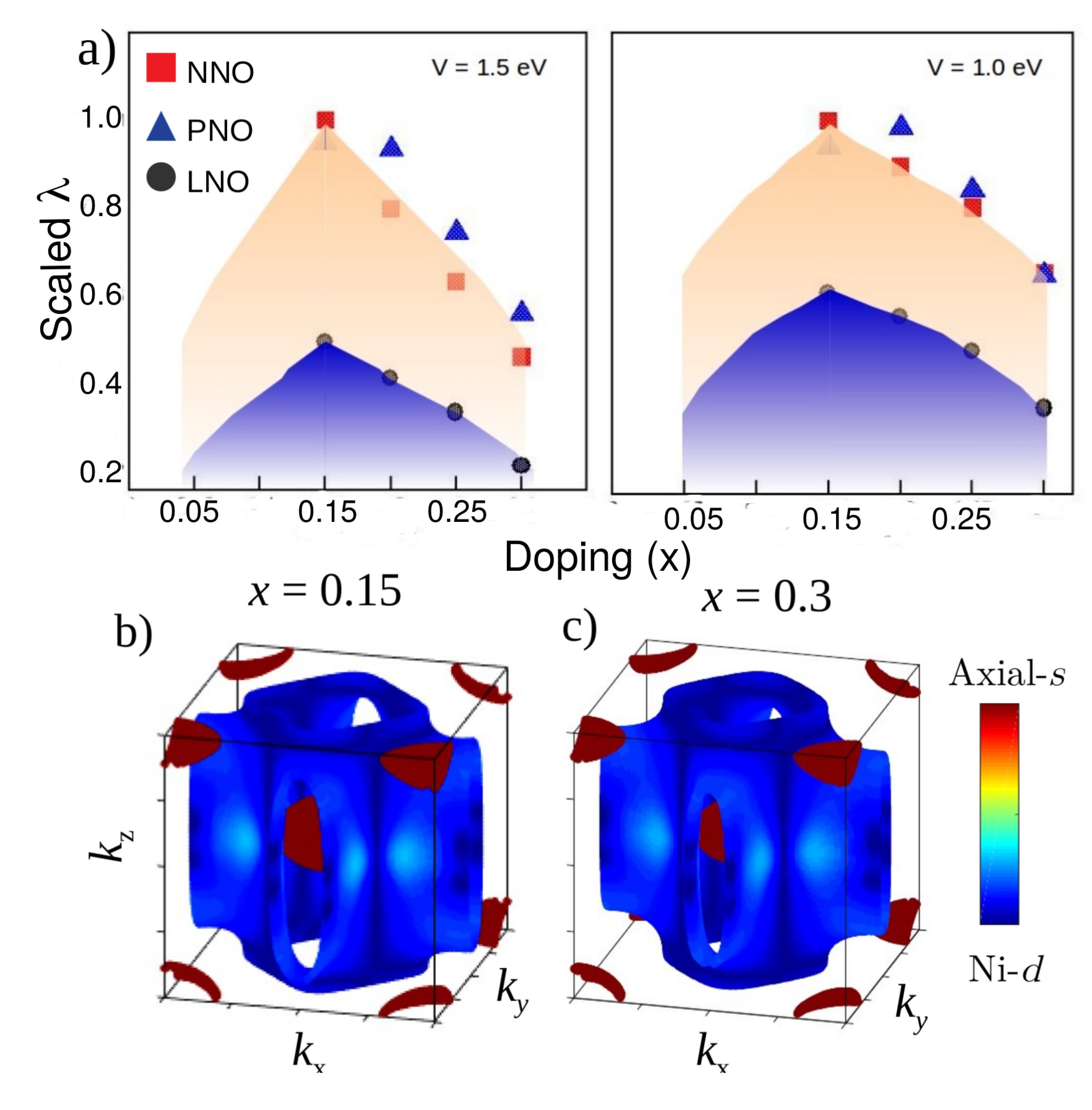}}
\end{center}
\caption{(Color online) a) Doping dependent SC coupling constant $\lambda$ (scaled by the strength of maximum $\lambda$ for NNO) for three different materials for choice of V = 1.5 eV and V = 1.0 eV. The shaded regions are the guide to the eyes. Due to the limitations of weak coupling theory,
  the shaded area for $x$ $<$ 0.15 are only schematic construction.
b-c) FS for NNO at two dopings, with the blue to red color map denoting the corresponding orbital weight for the Ni-$d_{x^2-y^2}$ to axial-$s$ orbitals.
}
\label{fig3}
\end{figure}

The highest pairing eigenvalue of Eq.~\eqref{SC2} gives a two-dimensional $d_{x^2-y^2}$ symmetry gap in the Ni-$d$ orbital channel. The $d_{x^2-y^2}$-wave state results from the antiferromagnetic ${\bf Q}=(\pi,\pi)$ nesting between the two hot-spots across the Brillouin zone (BZ) for the hole pocket (near the $k_z=0$ plane). The observed three-dimensionality of the hole-pocket clearly weakens the nesting strength and brings in doping dependence (see below). All three compounds have similar hole pocket and hence gives a $d_{x^2-y^2}$- wave solution. For NNO and PNO, there exists large area FS electron pockets $\alpha$ (centered around the $\Gamma$-point) which opens up another nesting channel between the $\alpha$ and $\gamma$ (centered around the $A$-point) electron pockets,
contributed by axial orbital, and offers an additional pairing channel which is of $d_{z^2}$-wave symmetry.
This is nearly absent in LNO due to diminishing presence of electron pocket at $\Gamma$. Thus
superconductivity in NNO and PNO is of two gap nature, while a single gap superconductivity is found in LNO. Furthermore, having a second pairing channel with a symmetry that is consistent with the corresponding FS nesting, enhances the SC coupling constant $\lambda$ in NNO and PNO, compared to LNO.  

The $d_{x^2-y^2}$ -wave gap gives a nodal quasi-particle DOS while the $d_{z^2}$ -wave gap symmetry becomes node-less as it originates from the axial orbital which primarily contributes in the $\alpha$ and $\gamma$ FS pockets, and has no contribution in $k_z=\pi/2$. This finding corroborates the recent STM data on NNO.\cite{STM}

In Fig. 3a), we plot the calculated values of $\lambda$ contributed by all channels, as a function of doping for the three compounds. The solution of gap equation shows that the pairing symmetry for NNO and PNO remains of two gap nature throughout the doping range, while that of LNO remains predominantly a single $d$-wave gap SC. The added
contribution of two channels of pairing in NNO and PNO, as opposed to a nearly
single channel SC in LNO, makes the pairing strength in 
about twice larger in NNO/PNO  compared to LNO in almost the entire doping
range.

Further, as seen in Fig. 3a),  $\lambda$ decreases monotonically with doping for all three compounds. Doping dependent calculations are carried out for two choices of $V$ = 1.5 eV and 1.0 eV with $U$=1, 0.5 eV for the Ni-$d$ and axial orbitals, respectively. We find that the qualitative features remain unchanged upon
change of $V$ value. Thus, the doping dependence arises purely due to changes in electronic structure, and guided by how the FS volume, FS nesting and the associated orbital weight evolve with doping. 

The origin of decreasing strength of SC with doping can be traced back to the FS area and nesting strength. As seen from the orbital resolved DOS in Fig.~1 (lower panel), with hole doping, the DOS of the Ni $d_{x^2-y^2}$ orbital increases, while that of the axial orbital decreases. There is a van-Hove singularity (VHS) of the $d_{x^2-y^2}$ which lies below the Fermi level, as in cuprates, however, this VHS cannot be doped within the experimentally feasible range. In a simple BCS like picture, one would thus expect the SC strength to increase with hole doping due to the increment of DOS. However, contrary to this, superconductivity is found to decrease with doping.

To find out the reason behind this, we probe the 3D FS topology and the orbital weight distributions for the representative case of NNO in Fig. 3b) and 3c) at two characteristic dopings ($x=0.15$ and $x=0.3$). The FS evolution between these two doings is monotonic, and there is no significant change in the FS topology across this doping range. The large FS, which is dominated by the Ni-$d$ orbital (blue color) has an interesting transition from the hole-like FS (as in underdoped cuprates) near $k_z=0$  to an electron-like FS (as in overdoped cuprates) near $k_z=\pi$ plane. The transition occurs close to the $k_z=\pi$ plane and this transition point moves towards the $k_z=0$ planes with increasing hole doping. In other words, the hole-pocket become more three-dimensional with increasing doping and this makes a difference in the FS nesting at ${\bf Q}=(\pi,\pi)$. As the area of the hole-like FS topology reduces with doping, the FS nesting strength at ${\bf Q}$ also gradually reduces. Thus, the SC strength decreases monotonically. On the other hand, for NNO/PNO the electron-pocket size decreases with hole doping and hence both its nesting strength and the contributions to $\lambda$ from its DOS decreases. In essence, the SC coupling constant for the $d_{z^2}$ also decreases with doping.

Our theory reproduces the right-hand side of the SC dome, as observed in NNO as well as in PNO, \cite{SCDome1,SCDome2}  The decrease of $\lambda$ at lower doping is not obtained in our model, since our theory  does not include the renormalization effects on the quasiparticle spectrum due to many-body interaction. It is known experimentally that the low-doping region is a correlated metal or weak insulator\cite{SCDome1,SCDome2}  and hence indicating the important role of correlation which quenches the SC coupling constant.\cite{Dolui,Das2domes} 

{\it Conclusion. -} Motivated by the recent discovery of superconductivity in NNO\cite{nno-sc} and in PNO\cite{pno} with very similar superconducting transition temperatures, we investigate
the role of non-zero $f$-ness of Nd and Pr compounds, as superconductivity has not
yet been observed in another infinite layer nickelate LaNiO$_2$, containing $f^0$ La.
While the active role $f$ electrons is debated, we discuss the role of $f$ electrons
in influencing the positions of 5$d$ levels of RE elements, thereby dictating the
nature of axial orbital contributing to the second band that crosses the Fermi level.
This provides subtle differences between the Fermi surface topology of the Pr and Nd compounds, and that of La compounds, driving the two gap superconductivity in Pr and Nd compounds as opposed to a single gap in La compound. Interestingly the doping dependent superconductivity shows a factor of two suppression in the strength of superconductivity
in La compound, as compared to that of Nd and Pr compounds.

Note, our calculation and analysis does not include of effect of magnetism of 4$f$ electrons. However, the fact that the difference and similarity of the three compounds have been brought out correctly, suggests the role of the $f$ electrons, as identified in the present study, to be the dominant one.

\acknowledgements
T.S-D acknowledges financial support from Department of Science and Technology, India. I.D acknowledges DST-TRC and SERB (Project No. EMR/2016/005925) for financial support. T.S-D and I.D acknowledge support from the National Science Foundation under Grant No. NSF PHY-1748958 and hospitality from KITP where part of this work was performed. TD acknowledges supports from the MHRD, Govt. of India under STARS research funding. The authors gratefully acknowledge the help of Anita Halder in making some of the figures.

$\ast$ Equal contribution.


\begin{thebibliography}{99}
\bibitem{suppl}  A plane wave cut-off of 600 eV was set to get good convergence of total energy and a k-mesh of $16\times16\times16$ was used for the Brilliouin zone integration. The energy convergence criterion was set to $10^{-6}$ eV during the energy minimization process of the self-consistent cycle.
\bibitem{note} The summation over orbital indices includes both intra- ($\Delta_{11/22}$) and inter- ($\Delta_{12/21}$) orbital pairing contributions. However, for the present case inter-orbital term is negligibly small.
\end{thebibliography}
\end{document}